\documentclass[useAMS,usenatbib,fleqn]{mnras}			
\usepackage{times}
\usepackage{natbib} 
\usepackage[]{graphicx,color}
\usepackage{amsmath}
\usepackage{amssymb}
\usepackage{textcomp}
\usepackage[normalem]{ulem}
\usepackage{rotating}

\title[Mapping the KDC in NGC~1407]
{Mapping the Kinematically Decoupled Core in NGC~1407  with MUSE}
\author[E. J. Johnston et al]{Evelyn~J.~Johnston$^{1, 2}$\thanks{Email: ejohnston@astro.puc.cl}, George K.~T. Hau$^1$, Lodovico Coccato$^{3}$  \& Cristian Herrera$^{1}$  \\
  $^1$European Southern Observatory, Alonso de C{\'o}rdova 3107, Casilla 19001, Santiago, Chile\\ 
  $^2$Institute of Astrophysics, Pontificia Universidad Cat{\'o}lica de Chile, Av. Vicu{\~n}a Mackenna 4860, 7820436 Macul, Santiago, Chile\\ 
  $^3$European Southern Observatory, Karl-Schwarzschild-Stra{\ss}e 2, 85748 Garching bei M{\"u}nchen, Germany 
}

\begin{document}

\maketitle

\begin{abstract}
Studies of the kinematics of NGC~1407 have revealed complex kinematical structure, consisting of the outer galaxy, an embedded disc within a radius of $\sim60$\arcsec, and a kinematically decoupled core (KDC) with a radius of less than 30\arcsec. However, the size of the KDC and the amplitude of the kinematic misalignment it induces have not yet been determined. In this paper, we explore the properties of the KDC using observations from the MUSE Integral Field Spectrograph to map out the kinematics in the central arcminute of NGC~1407. Velocity and kinemetry maps of the galaxy reveal a twist of $\sim$148\textdegree\ in the central $10$~arcseconds of the galaxy, and the higher-order moments of the kinematics reveal that within the same region, this slowly-rotating galaxy displays no net rotation. Analysis of the stellar populations  across the galaxy found no evidence of younger stellar populations in the region of the KDC, instead finding uniform age and super-solar $\alpha$-enhancement across the galaxy, and a smoothly decreasing metallicity gradient with radius. We therefore conclude that NGC~1407 contains a triaxial, kiloparsec-scale  KDC with distinct kinematics relative to the rest of the galaxy, and  which is likely to have formed through either a major merger or a series of minor mergers early in the lifetime of the galaxy. With a radius of $\sim$5~arcseconds or $\sim0.6$~kpc, NGC~1407 contains the smallest KDC mapped by MUSE to date in terms of both its physical and angular size.

\end{abstract}


\section{Introduction}\label{sec:introduction}
Distortions in the kinematics of galaxies are considered to be a signature of an interaction with another galaxy at some stage in their history, such as accretion of material or a satellite galaxy, or disruption due to a flyby with another galaxy \citep{Kormendy_1984, Hernquist_1991}. Kinematically decoupled cores, or KDCs, were first discovered in Early Type Galaxies (ETGs) by \citet{Efstathiou_1982} using long-slit spectroscopy to study the stellar kinematics, and many further detections of KDCs have been made using this technique \citep[e.g.][]{Bender_1988, Franx_1988}. However, the introduction of wide-field, high spatial resolution Integral Field Spectrographs (IFS) have allowed us to map out the central kinematics of galaxies for the first time \citep[e.g.][]{Emsellem_2004, McDermid_2006}. From these velocity maps, the amplitude of the kinematics misalignment could be measured, allowing KDCs to be defined as counter-rotating, co-rotating, or prolate relative to the host galaxy. 

Kinematic misalignments have been detected in around 10\% of ETGs \citep{Krajnovic_2011}, although in isolated ETGs the fraction has been seen to increase to 40\% \citep{Hau_2006}. Using data from the ATLAS$^{\text{3D}}$ survey \citep{Cappellari_2011}, \citet{ Krajnovic_2011} proposed a classification of kinematic misalignments, in which KDCs are defined as having a twist of $> 30$\textdegree\ in the kinematic position angle (PA$_{kin}$) between the two components, and where the $k_1$ coefficient, which represents the amplitude of the bulk rotation in the velocity, drops to 0 in the transition region, reflecting little or no net rotation there. Using this classification, they detected KDCs in around 7\% of ETGs, with the majority residing within slow rotators-- galaxies which show little or no net rotation  \citep{Emsellem_2007}. KDCs do not share the same properties, and \citet{McDermid_2006} has proposed that they fall into two categories. The first are kiloparsec scale KDCs, which are typically older than 8~Gyrs and generally reside within slow rotators, and the second type are compact KDCs, which are generally younger ($\sim$1~Gyr), exclusively found within fast rotators, and rotate around the same axis as the host galaxy. The differences in the stellar populations of these KDCs and of the kinematics of their hosts suggest that these two types of KDC formed through different mechanisms. The younger stellar populations of the compact KDCs enhance the luminosity of the nuclei of their host galaxies, making them easier to detect than the kiloparsec-scale KDCs \citep{McDermid_2006}.

NGC~1407 is an elliptical galaxy at the centre of the NGC~1407 or Eridanus-A group. Based on it's rotation, it has been classed as a slow-rotator according to \citet{Longo_1994}, \citet{Proctor_2009} and \citet{Arnold_2014}. Kinematics studies of this galaxy have been carried out using long-slit spectra and the ``Stellar Kinematics from Multiple Slits" (SKiMS) technique  of \citet{Norris_2008}, in which the kinematics are measured in a series of slit-spectra over the structure of the galaxy and maps created using methods such as kriging \citep{Matheron_1963, Cressie_1990}. With the variations in spatial coverage and density of measurements in each study, the combination of these studies have revealed a complex kinematical structure within NGC~1407. 

\begin{figure*}
%  \includegraphics[width=1\linewidth]{maps_V.ps}
%\subfloat{%
  \includegraphics[clip,width=0.7\linewidth]{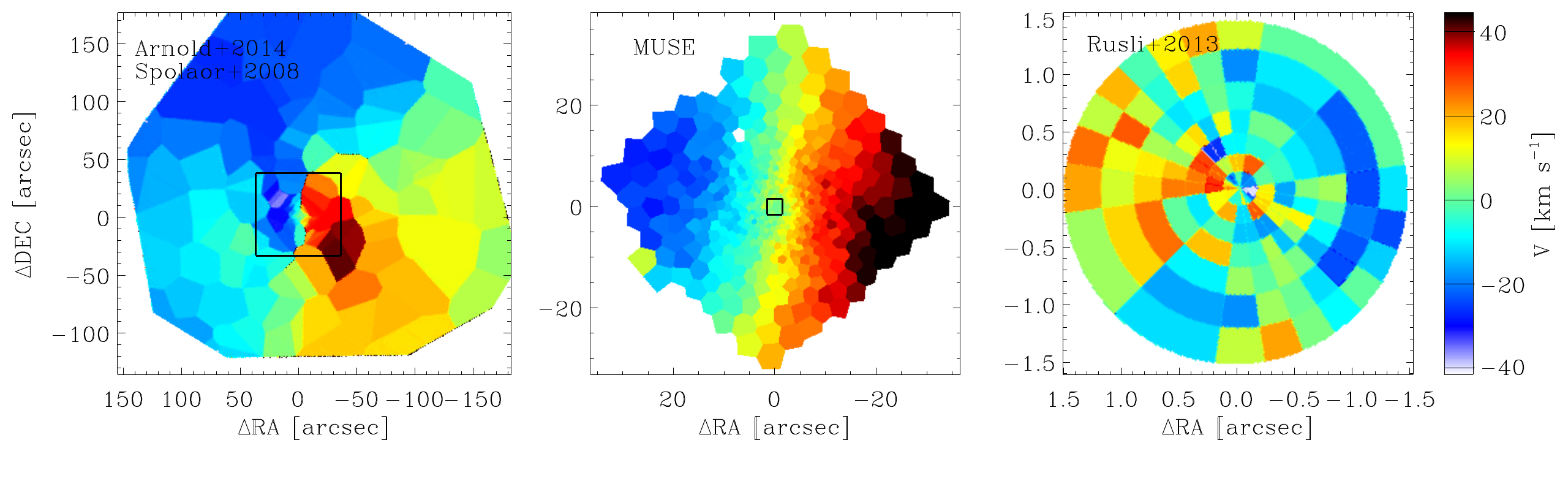}%
%}

%\subfloat{%
  \includegraphics[clip,width=0.7\linewidth]{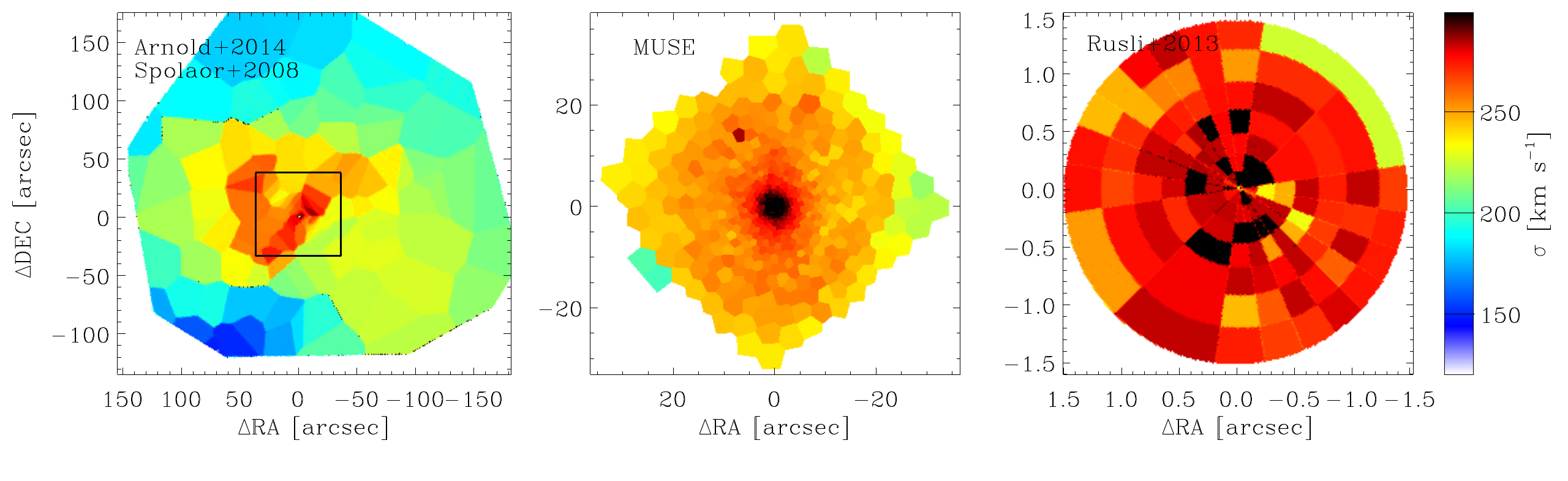}%
%}

%\subfloat{%
  \includegraphics[clip,width=0.7\linewidth]{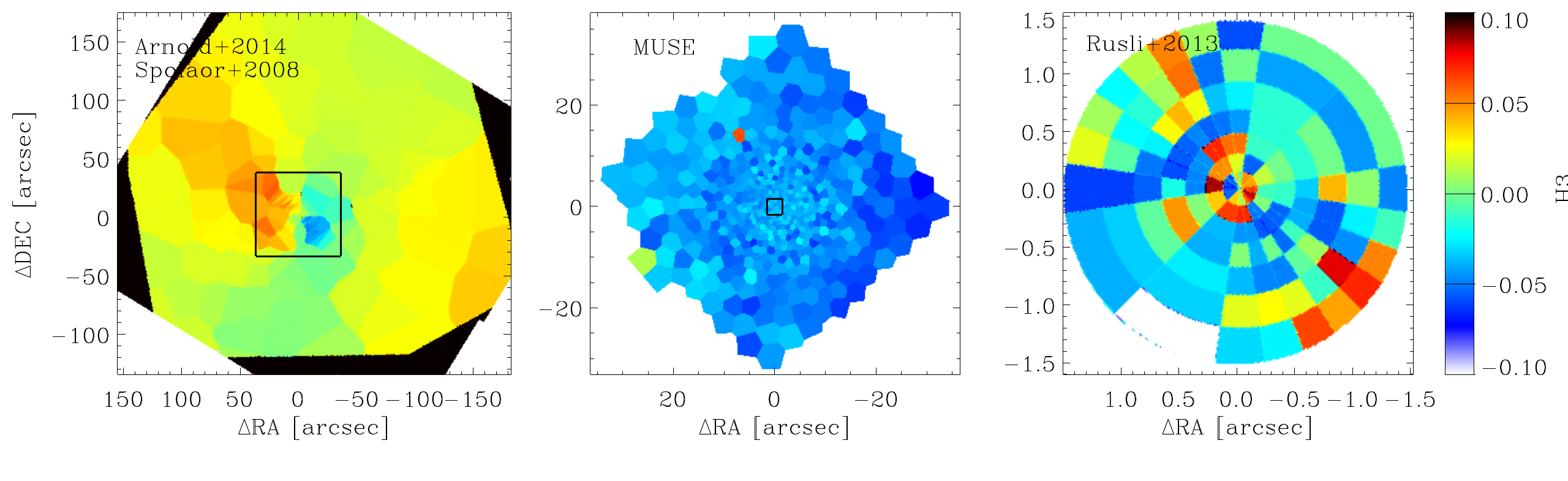}%
%}

%\subfloat{%
  \includegraphics[clip,width=0.7\linewidth]{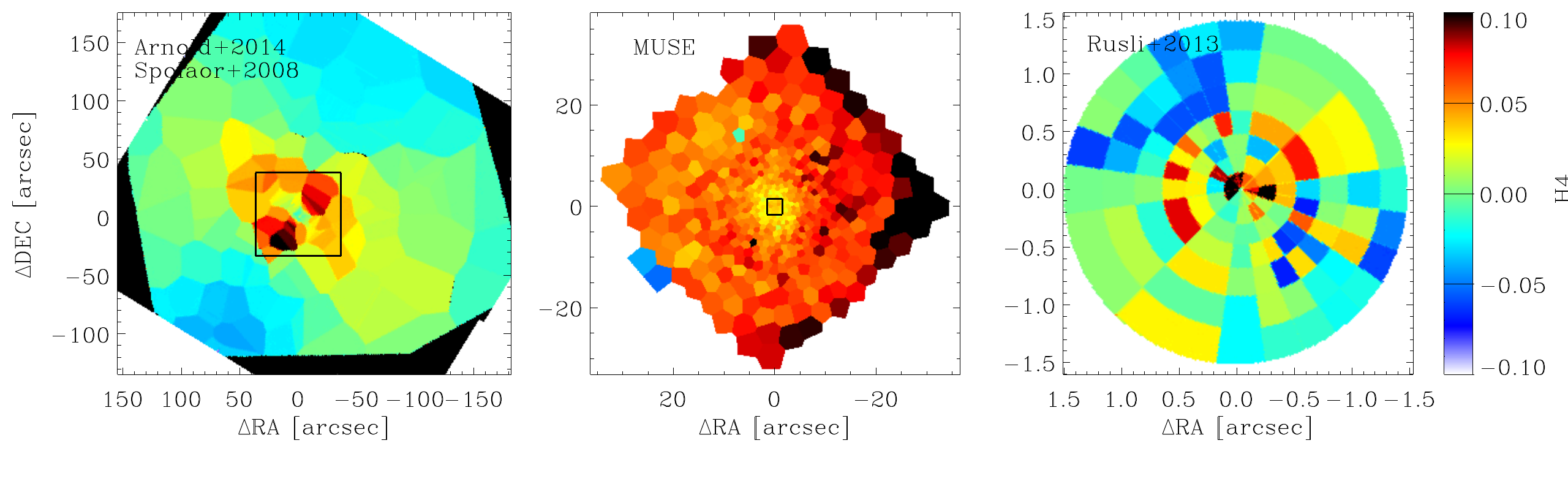}%
%}

  \caption{A comparison of the kinematics maps showing different kinematic components in NGC~1407. From top to bottom are the maps for velocity, velocity dispersion, $h_3$ and $h_4$. The plots in the left column are from  \citet[\textcopyright\ AAS, reproduced with permission]{Arnold_2014}, which used data from \citet{Spolaor_2008a} for the central region, and show the kinematics of the outer galaxy and the embedded disc within $\sim60$\arcsec. The middle column displays the MUSE data presented in this paper, and the right column shows the kinematics in the central region from \citet[\textcopyright\ AAS, reproduced with permission]{Rusli_2013}, which displays a third kinematic component that is misaligned relative to the embedded disc and outer galaxy. The black squares in the left and middle columns represent the fields of view of the middle and right columns respectively, and the colours in the plots along each row correspond to the colour bars on the right. All plots are orientated such that North is up and East is left.
 \label{fig:kinematics_comparison}}
\end{figure*}

The first kriging map of NGC~1407 was created by \citet{Proctor_2009} and covered radii between  $\sim30\!-\!180$\arcsec. They found a small variation in the PA$_{kin}$ of $\sim$10\textdegree\ at a radius of between $49\!-\!102$\arcsec.  More recently, \citet{Arnold_2014} and \citet{Foster_2016}  created more detailed kriging maps of NGC~1407  as part of the SAGES Legacy Unifying Globulars and GalaxieS survey \citep[SLUGGS, ][]{Brodie_2014}.  With an inner radius of 30\arcsec\ in the kinematics measurements, \citet{Foster_2016} found no significant distortion in the PA$_{kin}$. \citet{Arnold_2014} however combined the SLUGGS data with the long-slit measurements of \citet{Spolaor_2008a} to provide coverage of the inner regions of the galaxy. With this additional information, they detected  evidence of a rotating structure within 1R$_e$ (63\arcsec), which they believe is a signature of an embedded disc with kinematics that are misaligned from the photometric position angle. This distortion can be seen in Figure~\ref{fig:kinematics_comparison}, where the PA$_{kin}$ changes by about 60\textdegree\ inside and outside of 1R$_e$. This higher amplitude in the kinematic twist is likely due to the sparser sampling of \citet{Proctor_2009} giving a weaker result.

Studies of the inner regions of NGC~1407 have found evidence that the galaxy hosts a third kinematic component in addition to the embedded rotating disc within the outer rotating component.  Using long-slit spectra to measure the kinematics along the major-axis of this galaxy, \citet{Spolaor_2008a} first identified a possible kinematic twist in the line-of-sight velocity in the central 10~arcseconds of NGC~1407. While it's possible that this misalignment is due to the presence of a KDC distorting the kinematics, the authors admit that it may also be the result of the slit lying offset from the centre of the galaxy. A later study by \citet{Rusli_2013} mapped out the kinematics within the central 3~arcseconds of NGC~1407 using VLT/SINFONI in LGS mode to obtain diffraction-limited observations,  which revealed that this region of the galaxy rotates with a velocity of up to 40~km/s in the opposite direction to the rotation curve along the major axis obtained by \citet{Spolaor_2008a} and the inner embedded disc identified by \citet{Arnold_2014} (see Figure~\ref{fig:kinematics_comparison}). This result reflects the presence of a third component with distinct kinematics in the centre of NGC~1407 with a radius of between $1.5-30$\arcsec. For clarity, throughout this paper we will refer to the three kinematical components as the KDC, the embedded disc and the outer galaxy in order of their increasing size.

In order to determine the size of the KDC and the amplitude of the kinematic misalignment it induces relative to the kinematics of the embedded disc within which it resides, we need IFU data of this region with high enough spatial resolution to map out the twist in the kinematics. With a field of view of 1~arcmin$^2$ and a spatial resolution of 0.2~arcsec/pixel, the Multi-Unit Spectroscopic Explorer \citep[MUSE, ][]{Bacon_2010} at the Very Large Telescope (VLT) can map out the velocity profiles of the centres of nearby galaxies in enough detail to reveal small-scale kinematic distortions which may be hidden in data with poorer spatial resolution. MUSE has already proven useful for this type of analysis. For example, using MUSE Science Verification data, \citet{Emsellem_2014} confirmed the presence of a KDC of radius $15$~arcseconds ($\sim1.3$~kpc) at the centre of M87, with a kinematic twist of 140\textdegree\ and a velocity of amplitude $\pm$5~km/s. \citet{Krajnovic_2015} also used MUSE observations of NGC~5813 to confirm the presence of a counter-rotating core with a radius of 10~arcseconds ($\sim1.5$~kpc), and KDCs with prolate rotation (rotation around the long axis) have been detected in ETGs by  \citet{Iodice_2015} and \citet{Krajnovic_2018}.

In this paper, we  set out to analyse the kinematics and stellar populations of NGC~1407 using MUSE observations in order to measure the size and PA$_{kin}$ of the KDC. This paper is laid out as follows: Section~\ref{sec:DR} describes the data reduction, Sections~\ref{sec:Analysis} and \ref{sec:stellar_pops} outline the kinematics and stellar populations analyses respectively, and Section~\ref{sec:conclusions} discusses our conclusions. Throughout this paper we assume a Hubble Constant of $H_0=70.5$km~s$^{-1}$~Mpc$^{-1}$ \citep{Komatsu_2009}, giving a distance of 25.1~Mpc to NGC~1407 based on our measurement of the radial velocity of 1772$\pm$7~km~s$^{-1}$, where 1~arcsecond corresponds to 0.12~kpc.

%IFU_stellar_pops_radial_NGC1407

\begin{table*}
\centering
\begin{tabular}{cccccccc }
                        Data Set &  Date       & Prog. ID       &  PI  & Mode       & Exposure Time  &  No. Exposures & Mean Seeing      \\ 
\hline

\multicolumn{1}{l|} {$\cal A$} & {2014--12--15 }  & 094.B-0298        & Walcher     &  WFM-NoAO-N     & 145--150~s   &  10 &  0.80        \\
\multicolumn{1}{l|} {$\cal A$} & {2015--01--11 }  & 094.B-0298        & Walcher     &  WFM-NoAO-N     & 145--150~s   &  8 &  1.07        \\
\multicolumn{1}{l|} {$\cal B$} & {2015--09--07 }  & 095.B-0624        & Thomas     &  WFM-NoAO-N     & 900~s   &  2 & 1.94        \\

\end{tabular}
\caption{Details of the MUSE observations of NGC~1407 from the ESO Archive Facility.}
\label{Table:data}
\end{table*}

\section{Data Reduction}\label{sec:DR}

MUSE observations of NGC~1407 were found to be publicly available on the ESO Science Archive Facility. The galaxy was observed as part of two programs, the details of which are outlined in Table~\ref{Table:data}. Data set~{$\cal A$} covers a $1.8\times1.8$~arcminute mosaic centred on NGC~1407 while data set~{$\cal B$} consists of a single pointing in the centre of the galaxy. In both data sets, each pair of observations were rotated by 90~degrees and dithered slightly to reduce the appearance of the slicers and channels in the final combined datacube. Since the galaxy extends beyond the field of view of the MUSE data in both cases, separate sky exposures were observed for sky subtraction. Additionally, a standard star was observed on each night for flux and telluric calibrations, sky flats were taken during twilight within 7 days, and internal lamp flats were taken at the start of the observations to account for the time-dependent, temperature-related variations in the flux levels between each IFU. Due to the differences in the seeing for each set of observations, it was decided to reduce and analyse the two data sets independently for a better comparison.

The data were reduced using the ESO MUSE pipeline \citep{Weilbacher_2012} in the ESO Recipe Execution Tool (EsoRex) environment \citep{ESOREX}. A basic data reduction was carried out by creating master bias and flat field images and wavelength solutions for each detector, and applying these to the target, sky and standard star exposures. A flux calibration solution and sky continuum were obtained from the standard star and dedicated sky exposures, and applied to the corresponding science frames as part of the post-processing step. Finally, the reduced pixel tables created from the post-processing step for each exposure were combined to produce the final datacube. 

Data set~{$\cal A$} was observed over two nights though thin clouds with a single sky exposure taken per night. As a result, the sky background level was found to vary in each exposure, and the final mosaiced datacube showed an uneven sky subtraction across the field. In order to reduce this effect, the flux levels in overlapping regions of each pointing were measured in each exposure, and the sky continuum tables updated for each exposure to account for the differences in the flux relative to a reference exposure pointing at the centre of the galaxy. While this additional step reduced the artefacts in the background level across the mosaic, it did not remove them completely. However, for the kinematics analysis carried out in this study, an uneven sky background across the mosaic will not affect the results significantly. The analysis presented in this paper was carried out separately on both data sets and the results were found to be consistent. Unless otherwise specified, the figures presented in this paper show the results from data set~{$\cal B$}.

\section{Kinematics Analysis}\label{sec:Analysis}

\subsection{Stellar Kinematics}\label{sec:kinematics}

\begin{figure*}
\centering
\begin{minipage}{.5\textwidth}
  \centering
  \includegraphics[width=0.69\linewidth]{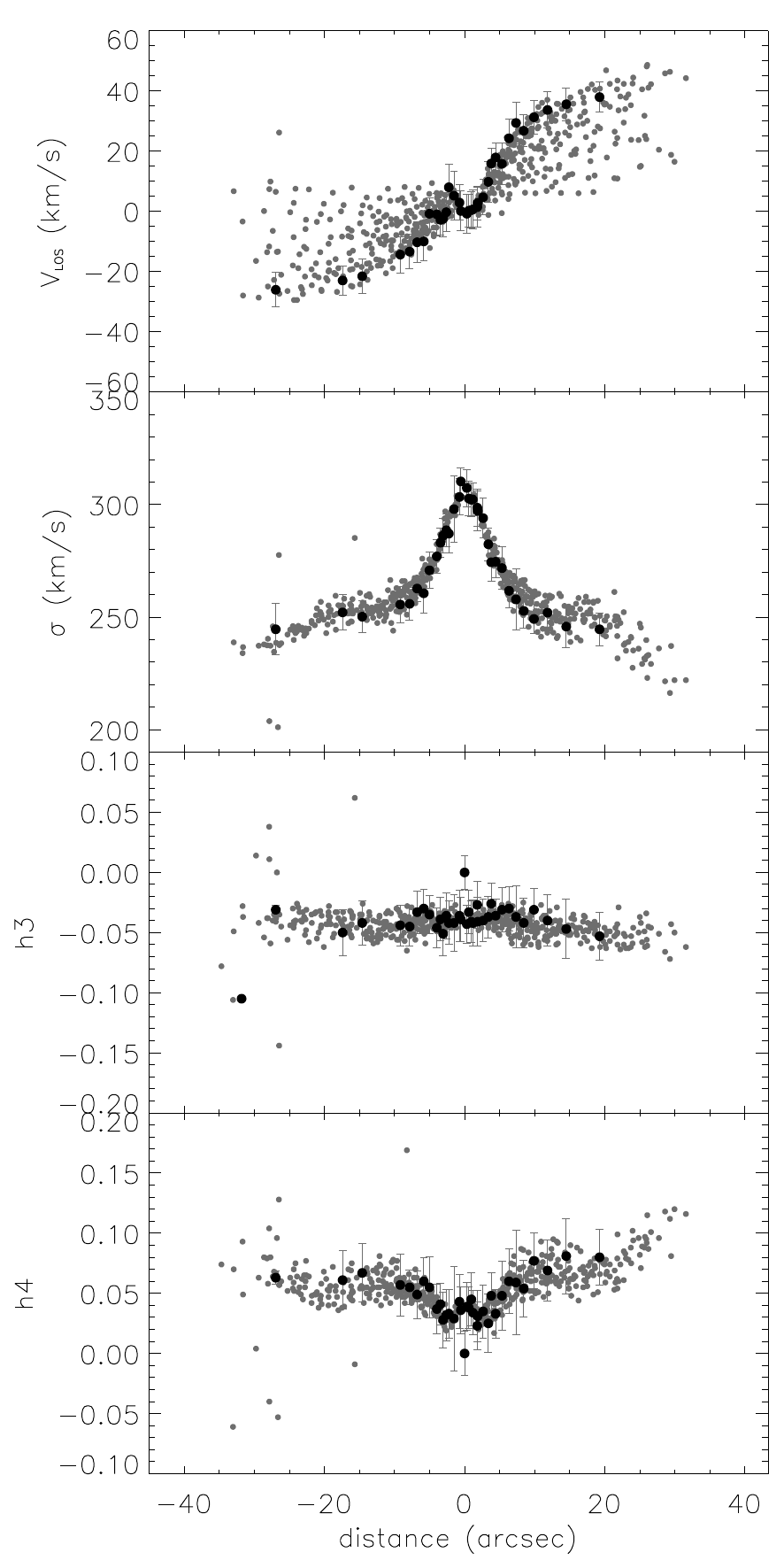}
\end{minipage}%
\begin{minipage}{.5\textwidth}
  \centering
  \includegraphics[width=1\linewidth]{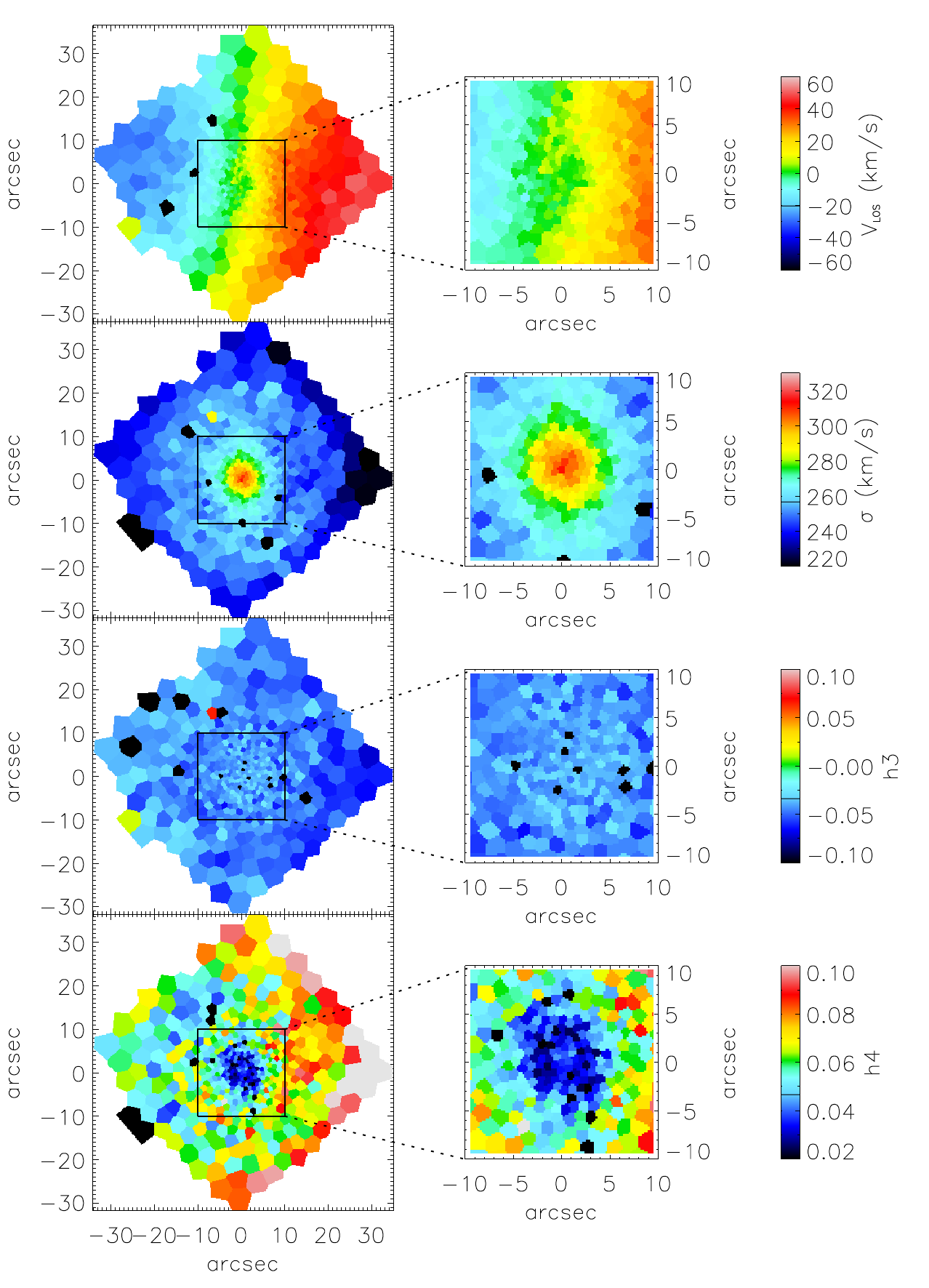}
\end{minipage}
 \caption{Kinematics measurements for NGC~1407 assuming one component is present, showing from top to bottom the line of sight velocity, velocity dispersion, h3 and h4 polynomials. On the left are the measurements for all Voronoi binned spectra plotted against the distance from the centre of the galaxy (grey points), with the measurements from those bins that fall within a slit of 1.2~arcsec along the major axis shown in black. On the right are the kinematics maps, showing both the full field-of-view of the MUSE data and a zoom-in of the central 20~arcsec. The maps are orientated with north to the top and east to the left.
\label{fig:kinematics_1}}
\end{figure*}
%kinematics_2comp_h.pro, NGC1407_deeper

The datacubes of NGC~1407 were first spatially binned using the Voronoi tessellation method of \citet{Cappellari_2003}, and the kinematics of the binned spectra were measured with the penalised Pixel Fitting (pPXF) code of \citet{Cappellari_2004}. pPXF combines a series of stellar template spectra of known ages and metallicities, and convolves them with a range of line-of-sight velocity distributions (V$_{\text{LOS}}$), velocity dispersions ($\sigma$) and  Gauss-Hermite polynomials ($h_3$ and $h_4$), to produce a model spectrum that best fits the binned galaxy spectrum and its kinematics. For this study, the MILES stellar library \citep{Sanchez_2006} was used, consisting of 156~template spectra ranging in metallicity from -1.71 to +0.22, and in age from 1 to 17.8~Gyrs. To achieve the best fit, the template spectra were multiplied by low-order Legendre polynomials to model out any mismatch in the flux calibration of the continuum. Figure~\ref{fig:kinematics_1} presents the kinematics results as measured between $4830-6200$\AA, both as maps and plots of the kinematics as a function of distance from the centre of the galaxy. For clarity in the plots, the measurements along the major axis of the galaxy are highlighted in black, having been selected by those bins whose centres fall within a long-slit of width 1.2~arcseconds (6 pixels) positioned over the centre of the galaxy with a position angle of 85\textdegree, where the centre of the galaxy was determined as the spaxel with the peak flux in the white-light image of the datacube. The uncertainties are shown for the measurements along the major axis, having been estimated through series of Monte Carlo simulations on the best fit spectrum for each bin with simulated levels of noise to match the original spectrum. 

The velocity map in Fig.~\ref{fig:kinematics_1} shows that the galaxy is rotating weakly within the MUSE field-of-view with a maximum velocity of $\sim$30~km/s along the line of sight. At the centre of the galaxy, a clear twist can be seen in the line-of-sight velocity, indicating the presence of a distortion in the kinematics there, with a velocity amplitude of $<$10~km/s. A comparison with the literature in Fig.~\ref{fig:kinematics_comparison} shows that the orientation of the KDC rotation is aligned similarly to the central velocity map of \citet{Rusli_2013}, while rotation detected outside of the KDC belongs to the embedded disc that \citet{Arnold_2014} detects within a radius of 60\arcsec. 

The velocity dispersion plot  displays a sharp rise in the region of the KDC, increasing from $\sim$250~km/s in the embedded disc to $310\pm9$~km/s at the core, thus indicating a region of increased disorder in the stellar orbits. This central measurement is in agreement with \citet[$\sigma=313\pm9$~km/s]{Spolaor_2008a}, \citet[$\sigma\sim290-310$~km/s]{Rusli_2013} and with the plots in \citet{Proctor_2009} where they re-analyse the data of \citet{Spolaor_2008a} and show a central velocity of $\sim$300~km/s, although their comparison table lists a maximum central velocity dispersion of $270\pm7$~km/s. Our own re-analysis of the \citet{Spolaor_2008a} data shows a peak of $299\pm10$~km/s in the central velocity dispersion. This discrepancy with the results of \citet{Proctor_2009} may have arisen through our use of a larger number of stellar templates, where they used 17 templates, and the offset in our measurements from the long-slit spectra and the IFU data could be due to a misalignment of the slit in \citet{Spolaor_2008a} or the higher spectral resolution of the MUSE data. The MUSE velocity dispersion map in Fig~\ref{fig:kinematics_comparison} shows that the central peak is relatively circular and symmetric, contrasting to the elongated shape in the maps of \citet{Arnold_2014} in the same figure. Since the maps of \citet{Arnold_2014} contains a combination of multi-slit spectra in the outer regions and long-slit spectra in the inner part, the distortion they find in the velocity dispersion map is therefore likely to be an artifact of the sampling of their data.

No obvious trends are seen in the $h_3$ map, while the $h_4$ map shows a drop within the region of the KDC with a small rise in the very centre of the galaxy. This parameter reflects the kurtosis in the spectral features, or the asymmetry from a Gaussian shape. In this case, the positive $h_4$ in the outer parts of the field-of-view indicate a bias towards radial orbits, while the drop in the $h_4$ at the centre reflects a bias towards tangential orbits. The change in the $h_4$ value at the boundary of the KDC therefore indicates a change in the orbital bias, and thus provides further evidence for a KDC with triaxial orbits. The elongated shape of the region where $h_4$ drops in Fig.~\ref{fig:kinematics_1} in the region of the KDC is reflected in the $h_4$ map of \citet{Arnold_2014} in Fig.~\ref{fig:kinematics_comparison} both over the scale of the KDC and of the embedded disc. However, the sparser sampling of their data combined with only having measurements along the major axis in the central arcminute of their maps may have led to an artificial exaggeration of the elongation in the $h_4$ maps. We therefore believe that the feature seen in Fig.~\ref{fig:kinematics_1} reflects the shape of the KDC.

If the difference in the kinematics and/or stellar populations are sufficiently large between the embedded disc and the KDC, it is possible to repeat the spectral fitting using a 2-component model. In this case, pPXF will create two model spectra representing the two structures with distinct kinematics that best fits the binned galaxy spectrum when combined. This technique has been very successful at disentangling the kinematics and stellar populations of extended counter- and co-rotating discs \citep[e.g.][]{Coccato_2011,Coccato_2013,Johnston_2013, Fabricius_2014} and kinematically decoupled cores of galaxies \citep[e.g.][]{McDermid_2006, Emsellem_2014, Krajnovic_2015}. The velocity maps of the central 3~arcseconds of NGC~1407 in \citet{Rusli_2013} show a clear counter-rotation in the PA$_{kin}$ in the region of the KDC compared to the outskirts of the maps presented here, and show that the KDC rotates with a velocity of up to 40~km/s. However, as shown in Fig.~\ref{fig:kinematics_1}, the amplitude of the velocity distortion in the centre of NGC~1407 is $\sim$10~km/s at most. MUSE has a spectral resolution of 2.74~\AA\ (69~km/s) at 5000~\AA\ and the data is both seeing-limited and more coarsely binned than the AO-corrected SINFONI data used by \citet{Rusli_2013}. As a result, when these 2-component fits with pPXF were attempted, both over the same wavelength range and over the Calcium Triplet, no conclusive results could be derived for the kinematics of the KDC.

\subsection{Kinemetry}\label{sec:kinemetry}
According to \citet{Krajnovic_2011}, a kinematically decoupled core is defined by a change in the PA$_{kin}$ of larger than 30\textdegree\ between two adjacent components and the $k_1$ coefficient of the kinematic moments dropping to 0 in the transition region. Both of these properties must be consistent across at least 2 consecutive measurements for each component. Additionally, a counter-rotating core is defined when the difference in the PA$_{kin}$ is 180\textdegree\ within the uncertainties. A comparison of the velocity maps of \citet{Arnold_2014} and \citet{Rusli_2013} in Fig.~\ref{fig:kinematics_comparison} suggest that the KDC is counter-rotating relative to the embedded disc within NGC~1407. However, the small field of view of SINFONI data, showing only the central 3~arcseconds, is insufficient to measure the extension of the KDC. Therefore, to determine the size and amplitude of the kinematic misalignment of the KDC, the kinemetry code of \citet{Krajnovic_2006} was used to measure the variation in the PA$_{kin}$  and the $k_1$ coefficient across the  inner regions of the galaxy.

\begin{figure*}
\centering
\begin{minipage}{.63\textwidth}
  \centering
  \includegraphics[width=0.95\linewidth]{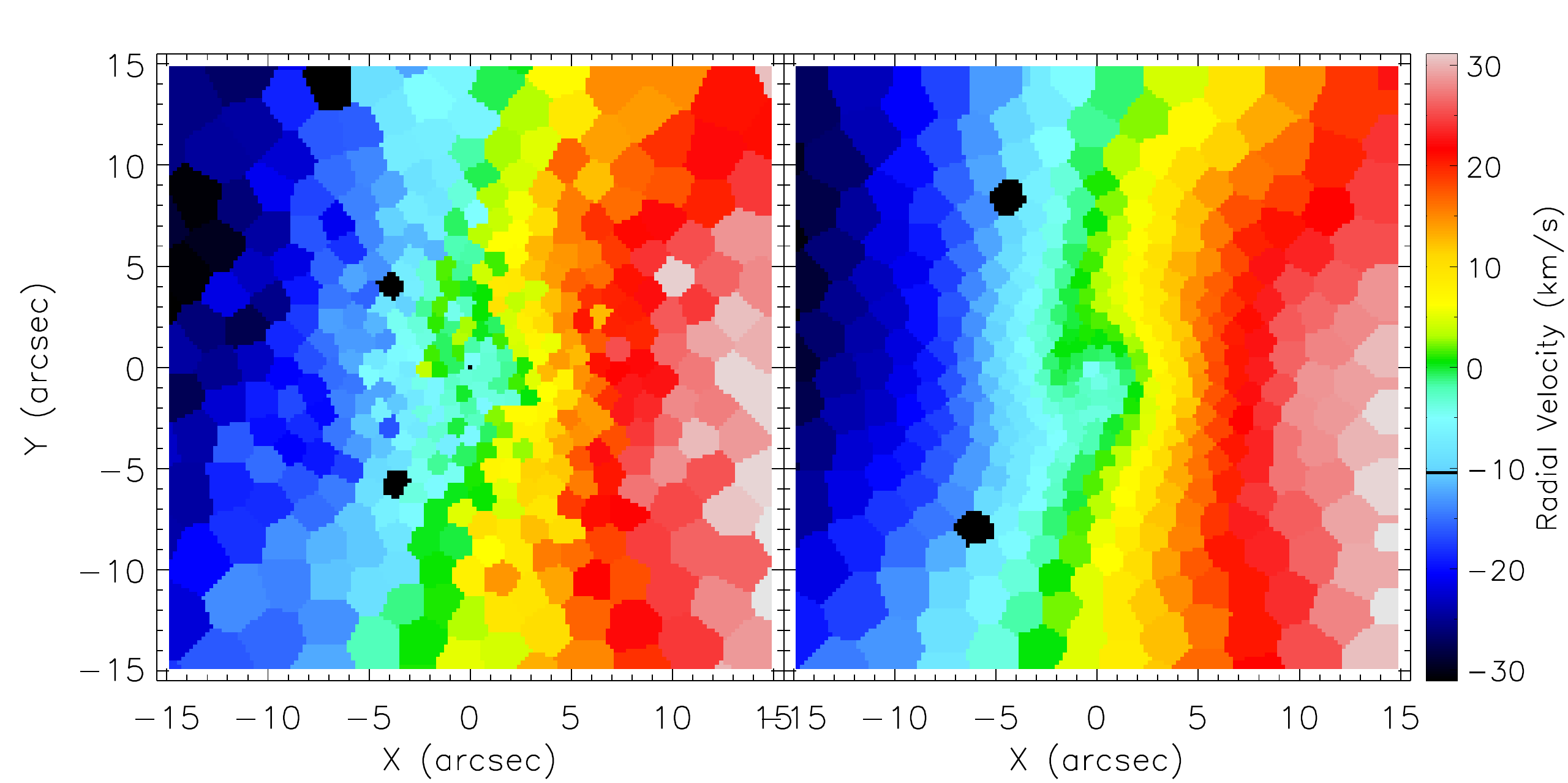}
   %\label{fig:kinemetry_maps}
\end{minipage}%
\begin{minipage}{.37\textwidth}
  \centering
  \includegraphics[width=0.95\linewidth]{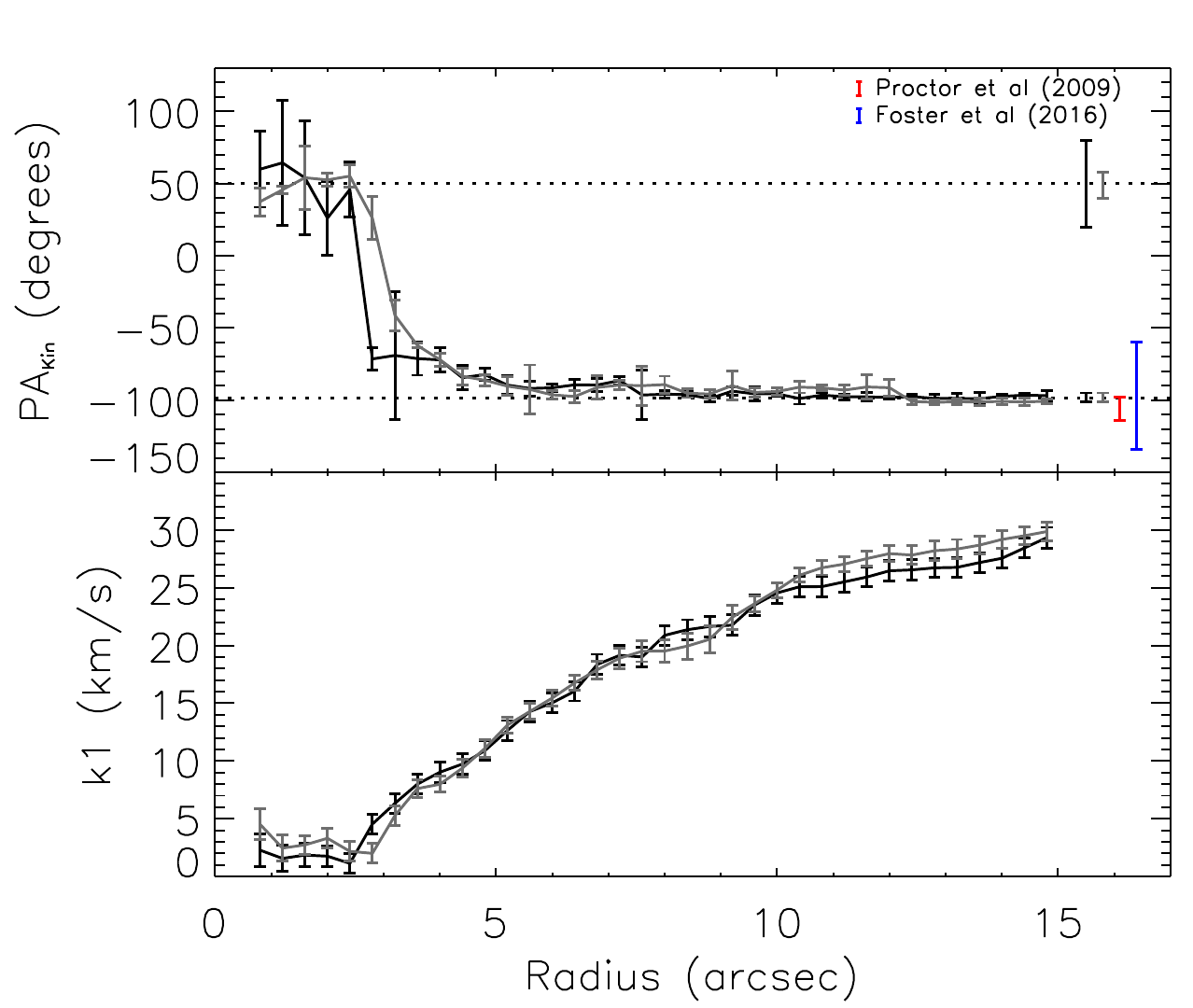}
   %\label{fig:kinemetry_plots}
\end{minipage}
 \caption{Kinemetry analysis for NGC~1407. Left: the input velocity map for the centre of the galaxy using data set~$\cal B$ alongside the smoothed kinemetric map of the same region. Right: the variation in PA$_{kin}$ (top) and the k$_1$ coefficient (bottom). Data set~$\cal A$ is marked in grey, and data set~$\cal B$  in black, and the error bars on the right represent the mean value for both datasets in grey and black, and the range of values measured for the embedded disc bu \citet{Proctor_2009} and \citet{Foster_2016} at radii between 30-130\arcsec. }
\label{fig:kinemetry}
\end{figure*}

NGC1407 is an unusually spherical galaxy- photometric decomposition of NGC~1407 by \citet{Huang_2013a} found that the galaxy was best modeled with a three-component fit, in which the effective radii of the three components were 5\arcsec, 22\arcsec and 124\arcsec, where all three components were found to have an ellipticity of 0.05. Similarly, \citet{Proctor_2009} measured the photometric axis ratio of NGC~1407 to be 0.95 using 2MASS data. Therefore, the kinemetry analysis was carried out with the axis ratio, $q$, constrained to be between 0.8 and 1, although the same trends were seen even without this constraint. The smooth reconstruction of the centre of the velocity map using up to the first-order kinemetric harmonic is shown in Fig.~\ref{fig:kinemetry} for data set~{$\cal B$}, alongside the same region of the original velocity map for comparison. A distortion is clearly seen in the smoothed map from the kinemetric analysis, and a comparison with the model maps in Fig.~1 of \citet{Krajnovic_2006} suggest that the velocity map in Fig.~\ref{fig:kinemetry} most closely resembles their models with a 90-180\textdegree\ change in PA$_{kin}$. 

Figure~\ref{fig:kinemetry}  also displays the plots for the variation in the PA$_{kin}$ and the $k_1$  coefficient measured from both data sets. Using data set~$\cal B$, the PA$_{kin}$ shows a twist of $\sim$148\textdegree, from 50~$\pm$~30\textdegree\ within a radius of 2.5~arcseconds to -98~$\pm$~3\textdegree\ outwards of a radius of 8~arcseconds. Similar results were obtained using  data set~$\cal A$--  a twist of $\sim$147\textdegree\ was detected, ranging from 49~$\pm$~9\textdegree\ to -98~$\pm$~3\textdegree. Due to the high spatial resolution and S/N of the MUSE data, the measurements for both the KDC and the host galaxy are consistent over at least five consecutive measurements in both data sets.

As shown in Fig.~\ref{fig:kinemetry}, the measurements at larger radii within this field are in agreement with the PA$_{kin}$ of \citet{Proctor_2009} and \citet{Foster_2016} for the embedded disc. \citet{Proctor_2009} measured a value of $254\pm8$\textdegree, or  $-106\pm8$\textdegree\ once  subtracted from $360$\textdegree\ to match the orientation used in this study, at a radius of 49~arcseconds, and   \citet{Foster_2016} measured values between 225 to 300\textdegree, or -135 to -60\textdegree, at radii between $30-130$~arcseconds.

%kinemetry_NGC1407_research_statement.pro, NGC1407_deeper (maps) NGC1407_deeper_int (plots), kinematics1. Reference plots- NGC1407_raw/kinemetry/kinematics

\begin{figure}
\centering
  \centering
  \includegraphics[width=0.8\linewidth]{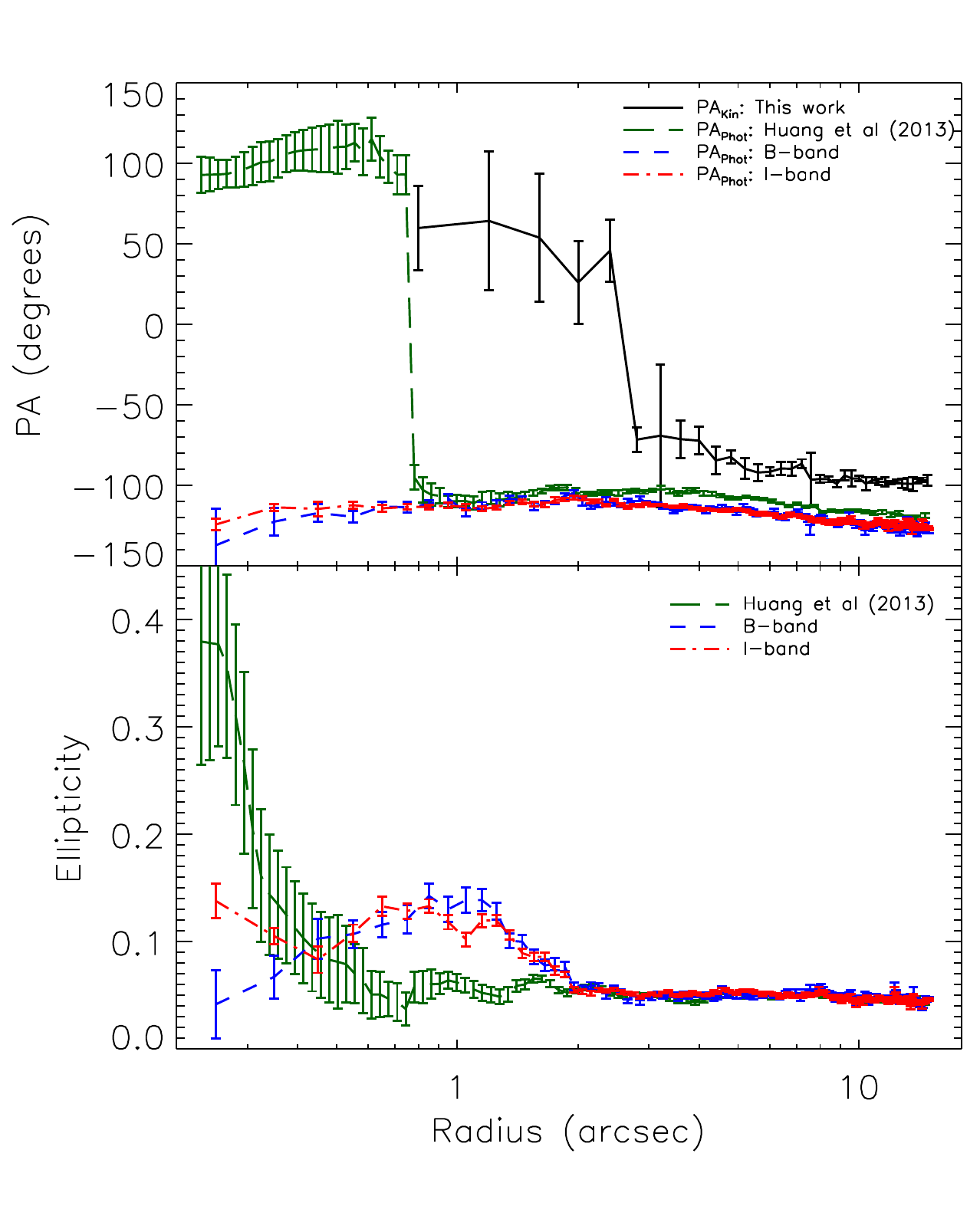}
   %\label{fig:kinemetry_plots}
 \caption{Top: A comparison between the PA$_{kin}$ from Fig.~\ref{fig:kinemetry} (black) and the PA$_{phot}$ from \citet{Huang_2013a} in the V-band (green) and with B- and I-band data from the HST/ACS (blue and red respectively). Bottom: A comparison of the ellipticities from the same three photometric data sets. The radius has been scaled logarithmically to better demonstrate the variations within the radius of the KDC.}
\label{fig:photometric_PA}
\end{figure}
%kinemetry_NGC1407_research_statement.pro, NGC1407_deeper (maps) NGC1407_deeper_int (plots), kinematics1. Reference plots- NGC1407_raw/kinemetry/kinematics

The $k_1$ coefficient represents the amplitude of the  rotation in the velocity of the model galaxy based on the velocity maps. Fig.~\ref{fig:kinemetry} reveals that $k_1$ follows a smooth curve which reflects the velocity curve in Fig.~\ref{fig:kinematics_1} outside of a radius of 3~arcseconds, while inside of this radius, it is flatter and close to 0, indicating that within this region  little net rotation can be detected with this data set. Since the determination of the kinematical centre of this galaxy is very subtle, we repeated the kinemetric analysis 900 times, each time centering on a different spaxel within a box of side 6\arcsec\ centered on the brightest spaxel in the white-light image. No significant rotation was detected in the region of the KDC in these fits, indicating that with the spatial and spectral resolution of the MUSE data available, we are unable to reproduce the amplitude of the rotation detected by \citet{Rusli_2013}.

With a kinematic twist of 148\textdegree, the two components are not quite counter-rotating with respect to one another. While the KDC clearly dominates the kinematics within a radius of $\sim$3~arcseconds, it's effects can be seen out to a radius of $\sim$5~arceconds in the velocity maps and the plot of the PA$_{kin}$. Assuming a distance of 25.1~Mpc and a scale of 0.12~kpc/arcsec on the datacubes, the KDC has a diameter of $\sim$1.2~kpc. This diameter rules out the possibility that the KDC is a nuclear stellar disc, which usually have diameters of a few tens to hundreds of parsecs in diameter \citep{Rest_2001}, and indicates that it is a larger embedded structure.

Since ellipticity of NGC~1407 is close unity, it is tricky to accurately determine the photometric position angle (PA$_{phot}$) of the galaxy to compare it to the PA$_{kin}$. A comparison of the radial variations in the PA$_{kin}$ and PA$_{phot}$ from \citet{Huang_2013a} is presented in Fig.~\ref{fig:photometric_PA}, with a 180\textdegree\ correction to match the orientation of the PA$_{kin}$, showing that outside of $\sim$8\arcsec\ the PA$_{phot}$ and the ellipticity are relatively constant. 
However, this data was observed at the Du~Pont 2.5m telescope with a seeing of 1.5\arcsec FWHM, meaning that any trends in the centre of the galaxy, in the location of the KDC, are likely to be smeared out. In order to compare the PA$_{kin}$ and PA$_{phot}$ in this region, imaging data of the system was obtained from the HST Legacy Archive for the F435W and F814W bands (B and I respectively) observed with the ACS/WFC. The PA$_{phot}$ and ellipticity were measured using the \textsc{iraf} task \textsc{ellipse} in the \textsc{stsdas/isophote} package. \textsc{ellipse} uses the methodology of \citet{Jedrzejewski_1987} to iteratively fit ellipses to the intensity of the galaxy to give a smoothed model. The results are plotted in Fig.~\ref{fig:photometric_PA}, and show  that both the HST data and the data from \citet{Huang_2013a} show a variation in the PA$_{phot}$ of about 20\textdegree\ between a radius of $\sim3$\arcsec, where the PA$_{kin}$ shows a strong transition in the stellar orbits, and the embedded disc. The lack of a strong variation in the PA$_{phot}$ to mimic the trend in the PA$_{kin}$ suggests that while the KDC dominates the kinematics in the centre of the galaxy, it does not dominate the light in that region. However, the variation in the ellipticity from the HST data within the central 4\arcsec of the galaxy reflects a slight elongation in this area. This trend was also detected in the same data set by \citet{Spolaor_2008a}. Interestingly, with a PA$_{phot}$ of $\sim-130$\textdegree, or $\sim50$\textdegree\ when corrected by 180\textdegree, this structure has a similar orientation to the PA$_{kin}$ of the KDC and the elongated $h_4$ profile in the centre of the galaxy shown in Fig.~\ref{fig:kinematics_1}. This similarity therefore suggests that the light from the KDC is not-negligible compared to the light at the centre of the embedded disc, and likely has an elongated structure which is reflected in the shape of the $h_4$ map.

Together, the kinematics and kinemetry results confirm the presence of a slightly elongated KDC present at the centre of NGC~1407 with a radius of $\sim$5~arcseconds, and which likely rotates slowly.

\section{Stellar Populations Across NGC~1407}\label{sec:stellar_pops}
The next question to consider is the formation of the KDC, which can be determined through analysis of the stellar populations within the KDC and the surrounding embedded disc. Events such as major mergers would lead to little difference in the stellar populations between the two kinematical structures since the stellar populations would be thoroughly mixed by the disruption in the stellar orbits \citep{Carollo_1997, Balcells_1998, Jesseit_2007, Bois_2010, Bois_2011}. On the other hand, a minor merger would produce a KDC that either retains the stellar populations of the progenitor accreted galaxy \citep{Kormendy_1984,Hernquist_1991, McDermid_2006}, or which shows enhanced metallicity, $\alpha$-enhancement and younger ages  due to the induced star formation following gas accretion \citep{Balcells_1998, Hau_1999}. An alternative process would be the creation of a rapidly-rotating KDC at the centre of an elliptical galaxy through in-situ star formation triggered by a flyby interaction with another  galaxy \citet{Hau_1994}.

In order to identify which of these scenarios was a likely formation mechanism for the KDC in NGC~1407, estimates of the stellar populations throughout the galaxy were obtained through analysis of the absorption line strengths.  To allow the line strengths across the galaxy to be displayed on the same model grids, the binned spectra used in Sections~\ref{sec:kinematics} and \ref{sec:kinemetry} were broadened to a FWHM of 14~\AA\ ($\sigma\sim349$km/s). The line strengths were then measured from the broadened spectra using the Lick/IDS index definitions \citep{Worthey_1994}, and the uncertainties were estimated from the propagation of random errors and the effect of uncertainties in the line-of-sight velocities \citep{Cardiel_1998}. 

\begin{figure}
  \includegraphics[width=1\linewidth]{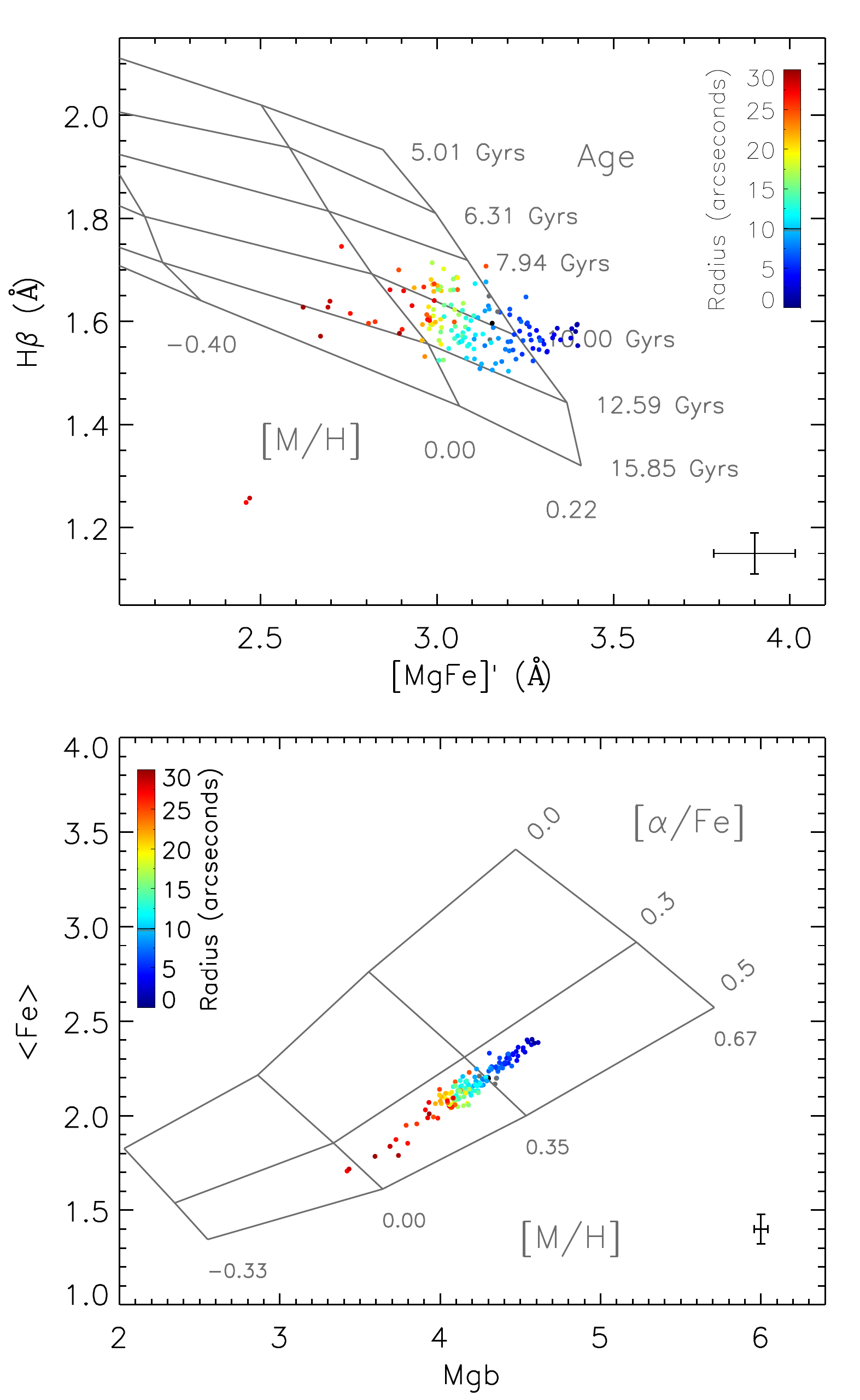}
  \caption{Line index measurements colour-coded according to distance form the centre of the galaxy, as shown by the colour bar. The top plot shows the age and metallicity estimates using the SSP models of \citet{Vazdekis_2010}, and the bottom plot shows the $\alpha$-enhancement using the models of \citet{Thomas_2011} for an age of 10~Gyrs. All data points and models were broadened to 14~\AA\ FWHM, and the error bars in the bottom right corners represent the mean uncertainty in the measurements.
 \label{fig:stellar_pops}}
\end{figure}

The  H$\beta$ feature and the combined metallicity index, [MgFe]$'$, were used as age and metallicity indicators respectively, where the latter was selected due to its negligible dependence on the $\alpha$-element abundance \citep{Gonzalez_1993,Thomas_2003}. To convert the line strengths into estimates of light-weighted age and metallicity, the measurements were over plotted onto the Single Stellar Population (SSP) models of \citet{Vazdekis_2010}. These models convolve the MILES stellar library \citep{Sanchez_2006}, which has a spectral resolution of $2.5\,$\AA\ (FWHM), with a Gaussian of the appropriate dispersion to reproduce the spectral resolution of the data. As a result, the SSP models are  matched to the data, minimizing the loss of information that normally occurs when degrading the data to match lower-resolution models. As shown in Fig.~\ref{fig:stellar_pops}, the galaxy is  uniformly old throughout, with an age of $\sim$10~Gyrs according to the models used here, with no significant age gradient, but it does show a negative metallicity gradient with radius. No obvious variations in the line strengths are visible within the region of the KDC. Application of full spectral fitting to a binned spectrum encompassing the full extent of the KDC also showed no evidence of multiple stellar populations that would suggest that the KDC is chemically distinct from the rest of the galaxy.

%IFU_stellar_pops_radial_NGC1407

The $\alpha$-element abundance across the galaxy can provide information about the star-formation timescale such that a higher $\alpha$-enhancement reflects shorter bursts of star formation activity. 
The $\alpha$-element enhancement across the galaxy was estimated by plotting the Mgb and $\langle\text{Fe}\rangle$ measurements onto the stellar population grids of \citet{Thomas_2011}  for solar-normalised logarithmic $[\alpha/{\rm Fe}]$ values of $0.0$, $0.3$ and $0.5$, with metallicities $[{\rm M}/{\rm H}]=\log(Z/Z_\odot)$ ranging between $-0.33$ and $+0.67$ for an age of $10\,$billion years. Since the models were measured at the Lick resolution, they were broadened to the Lick Index System (LIS) 14~\AA\ using the transformations of \citet{Vazdekis_2010}. The same metallicity gradient can again be seen in Fig.~\ref{fig:stellar_pops}, but no significant variation is seen in the $[\alpha/{\rm Fe}]$ values. NGC~1407 appears to be $\alpha$-enhanced throughout, with $[\alpha/{\rm Fe}] \sim0.35$, indicating that the galaxy and KDC were created in a rapid episode of star formation early in the lifetime of the galaxy. Due to the difficulties in distinguishing between stellar populations older than $\sim$10~billion years, we can only say for certain that the KDC and host galaxy formed over similarly short timescales early in the lifetime of the galaxy.

\section{Discussion and Conclusions}\label{sec:conclusions}
NGC~1407 is an elliptical galaxy at the centre of the NGC~1407 or Eridanus-A group that is known to have a complex kinematical structure. The galaxy contains an embedded rotating disc within 1R$_e$ (63\arcsec), which induces a distortion in the PA$_{kin}$ of about 60\textdegree\ at that radius \citep{Proctor_2009, Arnold_2014}, and inside that component resides a KDC with a radius $<30$\arcsec.

In this paper we have used MUSE data to map out the kinematics at the centre of NGC~1407 in order to measure the size of the KDC and the amplitude of the twist it induces on the line-of-sight velocity. The first step in the analysis was to bin the data and measure the kinematics across the field-of-view. The resultant velocity maps revealed a clear distortion at the centre of the galaxy, confirming the presence of a third component with distinct kinematics there. With the MUSE data used in this study, the measured amplitude of the velocity distortion in the region of the KDC was very low, of the order of 5-10km/s. This value is significantly lower than the $\sim$40km/s measured by \citet{Rusli_2013} in the central 3\arcsec of the galaxy, and this discrepancy is likely due be the poorer spatial resolution of the MUSE data combined with the effects of seeing.

The velocity dispersion map shows a peak in the region of the KDC, reflecting a region of less ordered stellar orbits and the presence of a black hole. According to \citet{Rusli_2013},  the black hole has a mass of $4.5\times10^9$M$_{\sun}$ and the diameter of its sphere of influence is 4.38\arcsec, putting it inside the KDC. This black hole may have had a role in the formation of the KDC. For example, simulations by \citet{Bekki_2000} have shown that the accretion of gas by the black hole could lead to the formation of a disc around it, and \citet{Holley_2000} found that the accretion of a dense stellar structure, such as an infalling galaxy, could lead to the formation of a counter-rotating core when a black hole is present.

Further evidence for the disruption of the stellar orbits lies in the $h_4$ maps for the MUSE data presented here and from \citet{Arnold_2014}, which display a drop within the region of the KDC. This trend indicates a transition from tangential orbits in the region of the KDC to radial orbits within the embedded disc. A similar $h_4$ profile was detected in the region of the KDC in NGC~5813 by \citet{Krajnovic_2015}.
Together, the kinematics maps presented here are reminiscent of those of NGC~4365 by \citet{vandenBosch_2008}, whose models for those maps represent the KDC as a triaxial structure at the centre of the galaxy. Therefore, it is likely that the KDC in NGC~1407 has a similar structure.

The next step in the analysis was to use the kinemetry code of \citet{Krajnovic_2006} to determine the amplitude in the distortion in the PA$_{kin}$ between the KDC and the embedded disc. This analysis revealed that the the PA$_{kin}$ of the KDC is offset by $148 \pm 31$\textdegree\ relative to the embedded disc, which is not quite consistent with counter-rotation between the two components. Furthermore, the radial variation in the $k_1$ coefficient revealed values close to zero within the region of the KDC which increase outside of a radius of 3\arcsec. Since the $k_1$ coefficient reflects of the amplitude of the rotation in the velocity based on the velocity maps, this trend indicates the presence of a zone of near-zero net rotation at the centre of the slowly rotating embedded disc. 

Based on the distortions in the kinematics maps and the kinemetry results, the KDC was found to have a diameter of $\sim$~1.2~kpc and to rotate with an amplitude of $\sim$5~km/s along the line of sight. The size rules out the possibility of the KDC being a nuclear stellar disc, which typically have diameters of a few tens to hundreds of parsecs \citep{Rest_2001}, and instead suggests that the KDC is a larger embedded structure at the centre of the galaxy which is likely to be slowly rotating based on the velocity maps of \citet{Rusli_2013}. This rotation however was not found in this study due to the smearing of the kinematics by seeing and the lower spatial resolution of MUSE data  compared to the seeing-corrected SINFONI data.

As a final step, the stellar populations were measured across the KDC and embedded disc to look for clues as to how the KDC formed. This analysis revealed that the both structures are uniformly old and $\alpha$-enhanced within the MUSE field of view, with a smooth negative metallicity gradient with radius from the centre of the galaxy. The majority of the literature agrees with the results presented here that no young stellar populations are present in the region of the KDC  \citep{Howell_2005, Thomas_2005, Humphrey_2006,Cenarro_2007, Zhang_2007,Spolaor_2008b}, with the only exception being \citet{Denicolo_2005}, who found evidence of a young stellar population in the nucleus with an age of $\sim$2.5~Gyrs. The uniformly old stellar populations reflect that the KDC formed a long time ago, around the same time as the host galaxy since there is no distinct signature in the age or metallicity at the boundary of the KDC, and the $\alpha$-enhancement indicates that the material in both the KDC and the embedded disc was formed over similarly short timescales early in the lifetime of the galaxy. While these results allow us to rule out the possibility of the KDC being formed through a recent accretion event or interaction with a passing galaxy,  the difficulties in distinguishing between stellar populations older than $\sim$10 billion years mean that we cannot determine if the KDC formed through a major merger or multiple minor mergers at high redshift.

We therefore conclude that the KDC is a triaxial structure that formed at around the same time as the host galaxy, which rotating slowly at an angle of 148\textdegree\ relative to the embedded disc within which it lies. The material was accreted into the galaxy through either a major merger or a series of minor mergers, which disrupted the stellar orbits and thoroughly mixed the stellar populations. The uniform old age of the stars indicate that the merger occurred long enough ago to allow evidence of any star-formation activity to fade. The combination of these old stellar populations and the size of the KDC are consistent with the kiloparsec-scale class of KDC defined by \citet{McDermid_2006}. With a diameter of $\sim1.2$~kpc, the KDC at the centre of NGC~1407 is the smallest KDC mapped to date with MUSE, both in terms of physical and angular size, and demonstrates the power of MUSE to resolve spectral signatures on such small scales.

\section*{Acknowledgements}

We would first like to thank the referee for their useful comments, which have improved this paper. We would also like to thank Caroline Foster for useful discussions that helped us understand the kinematical properties of this galaxy, Aaron Romanowsky \& Stephanie Rusli for allowing us to reproduce their figures in Figure~\ref{fig:kinematics_comparison} in this paper, and Song Huang for providing us with his PA and ellipticity data for NGC 1407 which has been used in Figure~\ref{fig:photometric_PA}. 
%E.J.J. acknowledges support from FONDECYT Postdoctoral Fellowship Project No. 3180557. 
This work was based on observations collected at the European Organisation for Astronomical Research in the Southern Hemisphere under ESO programmes 095.B-0624 (PI: Thomas) and 094.B-0298 (PI: Walcher), and obtained from the ESO Science Archive Facility, and observations made with the NASA/ESA Hubble Space Telescope, and observations obtained from the Hubble Legacy Archive, which is a collaboration between the Space Telescope Science Institute (STScI/NASA), the Space Telescope European Coordinating Facility (ST-ECF/ESA) and the Canadian Astronomy Data Centre (CADC/NRC/CSA).

%\bibliographystyle{mnras}
%
%\bibliography{paper6_refs}

\end{document}